\documentstyle[aps,prl]{revtex}
\draft

\def\mh{m_h^{}}
\def\gev{\rm GeV}
\def\fbi{\rm fb^{-1}}
\def\ww{W^*W^*}
\def\lsim{\mathrel{\raise.3ex\hbox{$<$\kern-.75em\lower1ex\hbox{$\sim$}}}}
\def\gsim{\mathrel{\raise.3ex\hbox{$>$\kern-.75em\lower1ex\hbox{$\sim$}}}}
\def\ljj{\ell\nu jj}
\def\ll{\ell\bar\nu \bar\ell \nu}
\def\etmiss{{\overlay{/}{E}}_T}
\def\ptmiss{{\overlay{/}{p}}_T}
\begin{document}

\twocolumn[\hsize\textwidth\columnwidth\hsize\csname
@twocolumnfalse\endcsname

\title {Extending the Higgs Boson Reach at the Upgraded Fermilab Tevatron}
\author{Tao Han and Ren-Jie Zhang}
\address{Department of Physics, University of Wisconsin, 
1150 University Avenue, Madison, WI 53706, USA}
\date{July, 1998}

\maketitle

\begin{abstract} 
We study the observability for a Standard Model-like Higgs boson
at an upgraded Tevatron via the modes
$p \bar p \to gg \to h \to \ww \to \ljj$ and $\ll$.
We find that with c. m. energy of 2 TeV and an integrated
luminosity of 30 fb$^{-1}$ 
the signal may be observable for the mass range of 
$135\ {\gev} \lsim \mh \lsim 180$ GeV at a $3-5\sigma$ statistical
level. We conclude that the upgraded Tevatron may have the potential
to detect a SM-like Higgs boson in the mass range from the LEP2
reach to 180 GeV.
\end{abstract}
]

The Higgs bosons are crucial ingredients in the Standard
Model (SM) and its supersymmetric extensions (SUSY).
Searching for Higgs bosons has been one of the major 
motivations in the current and future collider programs 
since they most faithfully characterize the mechanism 
for the electroweak gauge symmetry breaking. 
Experiments at LEP2 will eventually be
able to discover a SM-like Higgs boson with a mass about 
105 GeV \cite{lep2}. The LHC should be able to cover the full 
range of theoretical interest, up to about 1000 GeV \cite{lhc}.

It has been discussed extensively how much the Fermilab
Tevatron can do for the Higgs boson search. 
It appears that the most promising processes continuously 
going beyond the LEP2 reach would be the electroweak gauge
boson-Higgs associated production \cite{scott,GH,steve}
$
 p\bar p \to W h,\ Zh.
$
The leptonic decays of $W,Z$ provide a good trigger and 
$h\to b\bar b$ may be reconstructible with adequate $b$-tagging. 
It is now generally believed that
for an upgraded Tevatron with c. m. energy $\sqrt s=2$ TeV 
and an integrated luminosity ${\cal O}(10-30)\ \fbi$
a SM-like Higgs boson can be observed up to a
mass of about 120 GeV \cite{snowmass}. 
The Higgs discovery through these channels crucially depends 
up on the $b$-tagging efficiency and the $b\bar b$ mass resolution. 
It is also limited by statistics for $\mh > 120$ GeV. 
It might be possible to extend the mass reach to about
130 GeV via the decay mode $h\to \tau^+\tau^-$ \cite{steve}.
On the other hand, from the theoretical point of view, 
weakly coupled supersymmetric models generally predict the 
lightest Higgs boson to have a mass
$\mh\lsim 150$ GeV \cite{mhbound}. 
It would be of the greatest theoretical significance for the 
upgraded Tevatron to extend the Higgs boson coverage over
this range.

\begin{figure}[thb]
\vbox{\kern2.4in\includegraphics{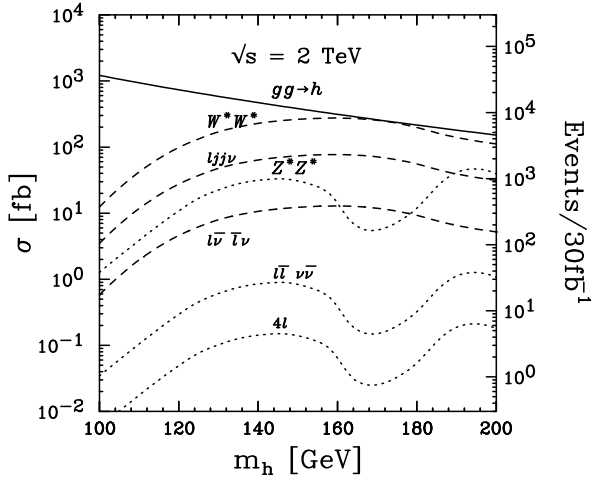}}
\caption{The Higgs-boson production cross-section via 
the gluon-fusion process versus $m_h$ at
the $2$ TeV Tevatron. The $h\to W^*W^*$ (dashes) 
and $Z^*Z^*$ (dots) 
channels and various subsequent decay modes are also depicted. 
}
\label{one}
\end{figure}

It is important to note that the leading production mechanism 
for a SM-like Higgs boson at the Tevatron is the gluon-fusion
process via heavy quark triangle loops. 
Although the decay 
mode $h\to b\bar b$ in this case would be swamped 
by the QCD background, $h\to W^*W^*$ mode
(where $W^*$ generically denotes a $W$ boson of either on-
or off-mass-shell)
will have an increasingly large branching fraction for
$m_h\gsim 130$ GeV and may have a chance to be observable. 
In this paper, we study in detail the observability of
a SM-like Higgs boson at an upgraded Tevatron for the modes
\begin{equation}
p \bar p \to gg \to h \to \ww \to \ell\nu jj\ \ {\rm and}\ \ 
\ell\bar\nu \bar\ell \nu ,
\label{sig}
\end{equation}
where $\ell=e,\mu$ and $j$ is a light quark jet.
In Fig.~\ref{one}, we show the cross section for $gg\to h $ 
versus $\mh$ with $p\bar p$ c.~m. energy $\sqrt s=2$ TeV. 
Along with the inclusive total cross section\footnote{
We have normalized our signal cross section to include
next-to-leading order QCD corrections \cite{spira},
and we use the CTEQ4M distribution functions \cite{cteq4m}.}
(solid curve), 
we show the $W^*W^*$ (dashes) and $Z^*Z^*$ (dots) channels,
as well as their various decay modes $W^*W^*\to \ljj,\ll$
and $Z^*Z^*\to \ell\bar \ell \nu\bar \nu, 4\ell$.
The scale on the right-hand side 
gives the number of events expected for
30 $\fbi$. We see that for the $\mh$ range of current
interest, there may be about 1000 events produced for
$\ww \to \ljj$ and about 100 events for $\ww \to \ll$.
This latter channel has been studied at the SSC and LHC energies
\cite{lnulnu} and at a 4 TeV Tevatron \cite{GH}. 
We find that at $\sqrt s=2$ TeV, 
after nontrivial optimization for the signal identification 
for Eq.~(\ref{sig}) over the substantial SM backgrounds,
it is promising to extend the Higgs boson reach at the
upgraded Tevatron with an integrated luminosity of 30 fb$^{-1}$ 
to $\mh \approx 135 - 180$ GeV at a $3-5\sigma$ statistically 
significant level.

\vskip 0.1in
\noindent
\underline{$\ww \to \ell \nu jj:$}

For this mode, we require the final state to have an isolated
charged lepton ($\ell$), large missing transverse 
energy ($\etmiss$), and two hard jets. 
The leading SM backgrounds are
\begin{eqnarray}
\label{QCD}
p\bar p &\to& W + 2\ {\rm QCD\ jets}, \ \
p\bar p \to W W\to  \ell \nu jj, \\
p\bar p &\to& W Z(\gamma^*)\to \ell \nu jj, \ \ 
p\bar p \to t \bar t\to  \ell \nu jj b\bar b .\nonumber
\end{eqnarray}
The background processes are calculated with the full 
SM matrix elements at tree level. 

To roughly simulate the detector effects, we use the following 
energy smearing
\begin{eqnarray}
\Delta E_j/E_j &=& 0.8/\sqrt E_j \oplus 0.05\quad {\rm for\ 
jets},\nonumber\\ 
\Delta E_\ell/E_\ell &=& 0.3/\sqrt E_\ell 
\oplus 0.01\quad {\rm for\ leptons},
\label{smear}
\end{eqnarray}
where $\oplus$ denotes a sum in quadrature. 
The basic acceptance cuts used here are
\begin{eqnarray}
&&p_{T\ell} > 15\ {\rm GeV},\ |\eta_\ell| < 1.1 ;~~~
p_{Tj} > 15\ {\rm GeV},\ |\eta_j| < 3 ;\nonumber\\
&&\etmiss > 15\ {\rm GeV};~~~\Delta R(\ell j) > 0.3,\ 
\Delta R(jj) > 0.7.
\label{basicjj}
\end{eqnarray}
For the sake of illustration, we 
present our study mostly for $\mh=140$ and 160 GeV and we will 
generalize the results to the full $\mh$ range of interest.
The cut efficiency for the signal is about $35\%$ (60\%) 
for $m_h\approx 140\ (160)$ GeV.

Since there are only two jets naturally appearing in the
signal events, the $t\bar t$ background can be effectively 
suppressed by rejecting events with extra hard jets. 
We therefore impose
\begin{equation}
{\rm Jet\ veto}\ \  p^{}_{Tj} > 15\ {\gev}\ {\rm in}\ |\eta_j| < 3.
\label{vetott}
\end{equation}
The QCD background in (\ref{QCD}) has the largest rate.
The di-jet in the signal is from a $W$ decay, while that in
the QCD background tends to be soft and collinear. We thus
impose the cuts on the di-jet:
\begin{eqnarray}
&&65 < m(jj) < 95\ {\rm GeV},\ \phi(jj) > 140^\circ; \nonumber\\
&&70 < m(jj) < 90\ {\rm GeV},\ \phi(jj) > 160^\circ,
\label{jjcut}
\end{eqnarray}
for $m_h=140$ and 160 GeV respectively,
where $m(jj)$ is the invariant mass of the di-jet and $\phi(jj)$
the opening angle of the two jets in the transverse plane.
For $m_h\geq 160$ GeV, the $m(jj)$ distribution has a unique 
peak because both $W$ bosons are on shell, so the $m(jj)$ cuts
in Eq.~(\ref{jjcut}) would not significantly 
harm the signal. On the other hand, 
for $m_h\leq 160$ GeV, nearly half of the
signal will be cut off by the $m(jj)$ cuts, making this
region of the Higgs mass more difficult to explore 
from this mode. 

We have also examined other mass variables, 
such as the $W$-boson transverse
mass $M_T(W)$, di-jet-lepton invariant mass $m(jj\ell)$,
and the cluster transverse mass $M_C$, which are defined as
\begin{eqnarray}
&&M_T(W) = \sqrt{(p_{T\ell}+\etmiss)^2
       -(\vec{p}_{T\ell} + \vec{\ptmiss}})^2,\nonumber\\
&&M_C = \sqrt{p_T^2(jj\ell) + m^2(jj\ell)} + \etmiss . 
\end{eqnarray}
The $M_T(W)$ develops a peak near $M_W^{}$ for on-shell $W$
decay. An upper cut on this variable below $M_W^{}$
can help remove the background 
from real $W$ decay as long as $\mh$ is less than 
$160$ GeV. 
The cluster transverse mass $M_C$ would 
be the most characteristic variable for the signal. 
It peaks near $\mh$ and yields a rather sharp
end-point above $\mh$.
To further improve the signal-to-background ratio $S/B$,
we find the following tighter cuts helpful
\begin{eqnarray}
&& 100<m(jj\ell)<120\ {\gev},\ \etmiss<30\ {\rm GeV},\  
120<M_C<140\ {\gev},\nonumber\\
&&\qquad\qquad 35<M_T(W) <55\ {\gev},\ 2.4<\Delta R(jj)<3.5; \nonumber\\
&& 100<m(jj\ell)<130\ {\gev},\ \etmiss<50\ {\rm GeV},\ 
130<M_C<170\ {\gev},\nonumber\\
&&\qquad\qquad 40<M_T(W) <90\ {\gev},\ 2.8<\Delta R(jj)<3.5,
\label{jjcut3}
\end{eqnarray}
for $\mh=140$ and 160 GeV respectively. 
 
We show the results progressively at different stages of the 
kinematical cuts in Table~\ref{tabI}. We see that for an
integrated luminosity of 30 fb$^{-1}$, the signal for $\mh=140$
GeV is very weak while that for $\mh\sim 160$ GeV can reach 
a 3$\sigma$ statistical significance.

\vskip 0.1in
\noindent
\underline{$W^*W^* \to \ell \bar \nu \bar \ell \nu:$}

For the pure leptonic channel, we identify the final state
signal as two isolated charged leptons and large missing
transverse energy. The leading SM background processes are
\begin{eqnarray}
p\bar p &\to& W^+W^-\to  \ell \bar \nu \bar \ell \nu,\ \ 
p\bar p \to Z Z(\gamma^*)\to \nu \bar \nu \ell \bar \ell, \nonumber\\
p\bar p &\to& t \bar t\to  \ell \bar \nu \bar \ell \nu b\bar b,
\ \  
p\bar p \to Z(\gamma^*)\to  \tau^+\tau^- \to 
\ell \bar \nu \bar \ell \nu \nu_\tau \bar \nu_\tau .
\label{dy}
\end{eqnarray}

We first impose basic acceptance cuts
\begin{eqnarray}
\nonumber
&&p^{}_{T\ell}  > 10\ {\gev},\ |\eta^{}_\ell| < 1.1;~~~
p^{}_{T\ell'} > 5 \ {\gev},\ |\eta^{}_{\ell'}| < 2.5;\\
&&m(\ell\ell')>10\ {\gev},~~~\etmiss > 25\ {\gev}.
\label{basic}
\end{eqnarray}
The cut efficiency for the signal is about 70\%. 
We also smear the lepton momenta according to Eq.~(\ref{smear}),
and veto the hard central jets via Eq.~(\ref{vetott}) to
effectively remove the $t\bar t$ background.
At this level, the largest background comes from the Drell-Yan
process for $\tau^+\tau^-$ production. However, this background
can be essentially eliminated by removing the back-to-back
lepton pair events by requiring 
\begin{equation}
\phi(\ell\ell) < 150^\circ.
\label{phi}
\end{equation}
The $\ww$ mass cannot be accurately reconstructed
due to the two undetectable neutrinos. However, both
the transverse mass $M_T$ and the cluster transverse mass
$M_C$, defined as
\begin{eqnarray}
&&M_T = 2\sqrt{ p^2_T(\ell\ell)+m^2(\ell\ell)},\nonumber\\ 
&&M_C =  \sqrt{ p^2_T(\ell\ell)+m^2(\ell\ell)}\ + \etmiss,
\end{eqnarray}
yield a broad peak near $\mh$.
We note that these transverse mass variables 
are very important for the signal identification and 
for controlling the systematic error. 

\begin{figure}[thb]
\vbox{\kern2.4in\includegraphics{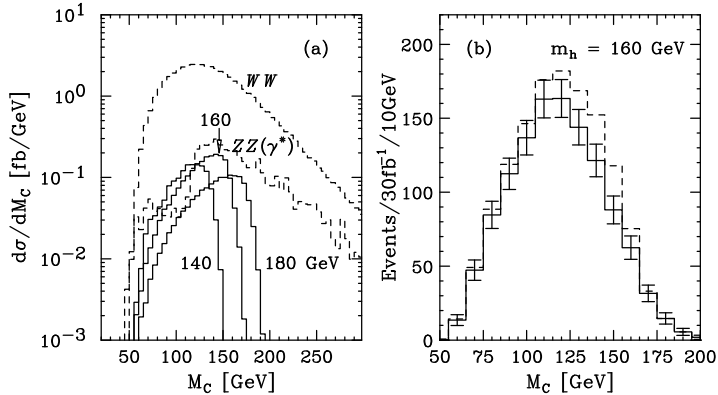}}
\caption{The cluster transverse mass distributions for 
$\ell{\bar\nu}{\bar\ell}\nu$ mode 
(a) for the signal $\mh=140,\ 160$ and 180 GeV
and the leading SM backgrounds 
with cuts (\ref{basic}) and (\ref{phi}).
With further selective cuts, we show the
event rates in (b) for SM background (solid)
and background plus signal (dashes) for $\mh=160$ GeV.
The statistical error bars are also indicated
on the background curve in (b).
\label{dist}}
\end{figure}

In Fig.~\ref{dist}(a), we show the $M_C$ distributions for 
the $\ll$ signal with $\mh=140,\ 160$ and 180 GeV along 
with the leading backgrounds after the cuts in 
(\ref{basic}) and (\ref{phi}). 
Although the mass peaks in $M_C$ are hopeful 
for signal identification, they are rather broad. 
We may have to rely on the knowledge of the SM 
background distribution. We hope that with the rather 
large statistics of the data sample, one may obtain good 
fit for the normalization of the background shape outside
the signal region in Fig.~\ref{dist}(a), 
so that the deviation from the predicted
background can be identified as signal.
Some additional useful cuts are 
\begin{eqnarray}
&&m(\ell\ell) < 80\ {\rm GeV}\ (70\ {\rm for}\ m_h \leq 140\ {\rm GeV}),
\nonumber\\
&&\etmiss < m_h/2, \ \  m_h/2 < M_C < m_h\ . 
\label{massc}
\end{eqnarray}
The results at different stages of kinematical cuts are shown in
Table~\ref{tabII}. Due to the absence of the large QCD background
in (\ref{QCD}), this pure leptonic mode seems to be statistically
more promising than the $\ell\nu jj$ mode. One may
expect a more than 3$\sigma$ (4$\sigma$) effect for $\mh=140$
(160) GeV with 30 fb$^{-1}$. 
It was pointed out in \cite{DD} that some angular variables
implement the information for decay lepton spin correlations
and are powerful in discriminating against the backgrounds.
With some further selective cuts on the angular distributions,
we find that the $S/B$ can be improved to about 8\% and 21\%,  
with the signal rates 2.6 and 3.3 fb for $m_h=140$ 
and 160 GeV, respectively. We show 
in Fig.~\ref{dist}(b) the event rate
distributions for the SM background
(solid) and signal plus background (dashes) for
$m_h=160$ GeV. The statistical 
error bars on the background are also indicated.

\begin{figure}[thb]
\vbox{\kern2.4in\includegraphics{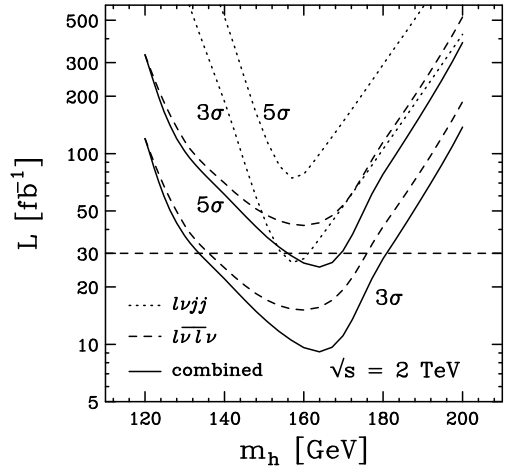}}
\caption{The integrated luminosity needed to reach $3\sigma$ and
$5\sigma$ statistical significance versus $m_h$. The dotted and 
dashed lines correspond to the $\ell\nu jj$ 
and $\ell{\bar\nu}{\bar\ell}\nu$ modes
respectively. The solid lines are the (quadratically) combined results.
\label{intL}}
\end{figure}

In Fig.~\ref{intL}, we show the integrated luminosities needed
to reach 3$\sigma$ and 5$\sigma$ significance versus $\mh$.
The dotted curves are for the $\ljj$ mode and the dashed for
$\ll$. We consider that these two modes have 
rather different systematic errors, 
so that we can combine the results for them quadratically. 
This is shown by the solid
curves. We see that with an integrated luminosity of 30 fb$^{-1}$,
one may be able to reach at least a 3$\sigma$ signal for 
$135\  {\gev} \lsim \mh \lsim 180\ {\gev}$. 
Taking into account the previous studies \cite{scott,steve,snowmass},
we conclude that the upgraded Tevatron with $\sqrt s=2$ TeV
and 30 fb$^{-1}$ may have the potential to detect the 
SM-like Higgs boson in the mass range from the LEP2 reach
to 180 GeV. On the other hand,
if there is only about 10 fb$^{-1}$ data available,
the sensitivity to the Higgs boson search via the modes
of Eq.~(\ref{sig}) would be very limited. A higher luminosity
is strongly called for in this regard.

Finally, a few remarks are in order.
(a) In our analyses, we have not considered
the $W \to \tau \nu_\tau$ mode.
Including this decay channel
would increase the signal rate by a factor of 3/2
(9/4) for $\ljj$ ($\ll$) mode. But the signal
identification would be more challenging.
The other modes such as $Z^*Z^* \to \ell \bar \ell jj,
\ell\bar \ell \nu \bar \nu$ and $4\ell$,
although smaller, may also be
helpful to improve the signal observability.
(b) Our results presented here are valid not
only for the SM Higgs boson, but also for
SM-like ones such as the lightest Higgs boson
in SUSY at the decoupling limit.
If there is an enhancement from new physics for 
$\Gamma(h\to gg)\times BR(h\to WW,ZZ)$ over 
the SM expectation, the signal of Eq.~(\ref{sig})
would be more viable.
If $BR(h\to b\bar b)$ is suppressed, such as in 
certain parameter region in SUSY, then the 
signal under discussion may 
complement the $Wh,Zh\ (h\to b\bar b)$ channels
at a lower $m_h$ region.

Our results summarized in Fig.~\ref{intL} 
based on the parton-level simulation are 
clearly encouraging to significantly extend the
reach for the Higgs boson search at the upgraded
Tevatron. The more comprehensive results with full Monte Carlo simulations
in a realistic environment will be reported
elsewhere \cite{future}.

{\it Acknowledgments}: We would like to thank V. Barger, 
E. Berger, R. Demina, M. Drees, T. Kamon, S. Mrenna,
J.-M. Qian, S. Willenbrock and J. Womersley for helpful 
comments. T.H. would like to thank the Aspen Center for
Physics for its hospitality during the final stage of the 
project. This work was supported in part by a DOE grant
No. DE-FG02-95ER40896 and in part by the Wisconsin Alumni Research 
Foundation.

\onecolumn{
\begin{table}[thb]
\begin{tabular}{|l|c|c|c|}
$\sigma$ [fb]  & Basic Cuts in (\ref{basicjj})
& Cuts in (\ref{jjcut}) & 
Cuts in (\ref{jjcut3}) \\
\hline
 $\mh$ [GeV] & 140 \ \ \ 160   & 140\ \ \ 160  &140\ \ \ 160 \\
\hline
 signal  &  23\ \ \ 49   & 8.8\ \ \ 21  &  2.2\ \ \ 15 \\
\hline
background     &     &  &        \\
$Wjj$     & $6.5\times 10^5$    & $6.1\times 10^3$\ \ \ $2.1\times 10^3$
    & $1.4\times 10^2$\ \ \ $5.5\times 10^2$  \\
$WW$          & $1.3\times 10^3$      
              & $5.5\times 10^2$\ \ \ $2.4\times 10^2$  
              & $17\ \ \ $55  \\
$WZ$            & 66      & 20\ \ \ 7.0  & 0.3\ \ \ 1.3   \\
$t{\overline t}$  & 0.4 & 0.1 &    0.0   \\
\hline
$S/B$ & -      & 0.1\%\ \ 0.9\%  &  1.4\%\ \ 2.5\% \\
$S/\sqrt{B}$ (30 fb$^{-1}$) & -  & 0.6\ \ \ 2.4  &  1.0\ \ \ 3.3    
\end{tabular}
\caption{$h\rightarrow \ww \rightarrow \ell\nu jj$ signal 
and background cross sections (in fb) for $\mh=140$ and 160 GeV, 
after different stages of kinematical cuts. 
A jet-veto cut in Eq.~(\ref{vetott})
has been implemented for the $t\bar t$ background.}
\label{tabI}
\end{table}

\begin{table}[thb]
\begin{tabular}{|l|c|c|c|c|}
$\sigma$ [fb]  & Basic Cuts in (\ref{basic})& $\phi(\ell\ell)<150^\circ$ & 
Cuts in (\ref{massc}) & Refined Cuts\\
\hline
 $\mh$ [GeV]& 140 \ \ \ 160   & 140\ \ \ 160  &140\ \ \ 160  &140\ \ \ 160  
\\
\hline
  signal   & 7.3\ \ \ 10 & 7.0\ \ \ 9.9  &  6.3\ \ \ 9.1  & 2.6\ \ \ 3.3 \\
\hline
background       &           &      &     & \\
$WW$            & $2.7\times 10^2$       & $2.4\times 10^2$ 
                & $1.1\times 10^2$\ \ \ $1.4\times 10^2$  &  32\ \ \ 16  \\
$ZZ(\gamma^*)$ & 24        & 18 &  1.8\ \ 1.8   &  0.3\ \ \ 0.1\\
$Z(\gamma^*)$    & $3.9\times 10^2$       & 0.0 &  0.0  & 0.0\\
$t{\overline t}$ & 0.2 & 0.1 &  0.0 &  0.0\\
\hline
$S/B$  & 1.1\%\ \ \ 1.5\% & 2.7\%\ \ \ 3.9\%  &  5.6\%\ \ \ 6.4\% 
                                              & 8.0\%\ \ \ 21\%\\
$S/\sqrt{B}$ (30 fb$^{-1}$) &1.5\ \ \ 2.1   & 2.4\ \ \ 3.4  &  3.3\ \ \ 4.2 
                                                            &  2.5\ \ \ 4.5
\end{tabular}
\caption{$h\rightarrow W^*W^*\rightarrow \ell\bar\nu
\bar\ell\nu$ signal and background cross sections (in fb)
for $\mh=140$ and 160 GeV, after different stages of 
kinematical cuts. The last column corresponds to the refinement of
mass cuts and various angular distribution cuts.
A jet-veto cut in Eq.~(\ref{vetott})
has been implemented for the $t\bar t$ background.}
\label{tabII}
\end{table}
}

\end{document}